# Hierarchical self-organization of non-cooperating individuals

T. Nepusz[a] and T. Vicsek[a,b]

[a]Department of Biological Physics, Eötvös University, Pázmány P. stny 1/A, 1117 Budapest, Hungary, and [b]Statistical and Biological Physics Research Group of the Hungarian Academy of Sciences, Pázmány P. stny 1/A, 1117 Budapest, Hungary

**Hierarchy is one of the most conspicuous features of numerous natural, technological and social systems[1,2,3,4]. The underlying structures are typically complex and their most relevant organizational principle is the ordering of the ties among the units they are made of according to a network displaying hierarchical features[5,6,7,8]. In spite of the abundant presence of hierarchy no quantitative theoretical interpretation of the origins of a multi-level, knowledge-based social network exists. Here we introduce an approach which is capable of reproducing the emergence of a multi-levelled network structure based on the plausible assumption that the individuals (representing the nodes of the network) can make the right estimate about the state of their changing environment to a varying degree. Our model accounts for a fundamental feature of knowledge-based organizations: the less capable individuals tend to follow those who are better at solving the problems they all face. We find that relatively simple rules lead to hierarchical self-organization and the specific structures we obtain possess the two, perhaps most important features of complex systems: a simultaneous presence of adaptability and stability. In addition, the performance (success score) of the emerging networks is significantly higher than the average expected score of the individuals without letting them copy the decisions of the others. The results of our calculations are in agreement with a related experiment[9,10] and can be useful from the point of designing the optimal conditions for constructing a given complex social structure as well as understanding the hierarchical organization of such biological structures of major importance as the regulatory pathways[11] or the dynamics of neural networks.[12]**

We have developed an approach to address the question of the spontaneous emergence of hierarchical networks, since in spite of the widespread presence of hierarchy most of the related questions are still open. Why is hierarchy so common? In the case of living matter or social systems there must be an advantage of such an organization, because of the permanent evolution of these systems preferring the more efficient (having a larger fitness) variants. But where is this advantage? Better adaptability? A more efficient, robust, stable structure? A faster spreading of relevant information? Remarkably, it is rather difficult to provide answers to the above questions in the context of a quantitative analysis. Prior attempts have shown that hierarchical organization can be advantageous, but either have not addressed the network aspect of the dominance[13,14] hierarchies or considered the embedded[15,16,17,18] cases and, consequently, have not resulted in explaining the emergence of the kind of multi-level hierarchical networks containing directed edges, i.e., flow-hierarchies, into which individuals in various societies (including both human and animal) are typically situated.

Building and investigating a simple model that spontaneously leads to hierarchical organization may significantly deepen our insight into collective decision making processes based on well structured leadership relations resulting in better performance. The most common approach used to treat complex social relations and the associated dilemmas has been network and game theory[19,20]. It has been widely acknowledged that relatively simple models (sets of rules) can adequately account for a number of social or economical situations. A number of outstanding results have been obtained to describe e.g., the behaviour of markets[21,22], while in another, quickly growing alternative direction, the so called agents interact in such a way that according to the corresponding rules of the models the agents are optimizing their benefits[23]. Consequently, we have designed a model, which due to its simplicity is applicable to a wide range of actual systems displaying hierarchical structure.

Thus, we look at a society as a system that evolves in the direction of optimizing the (not yet defined) benefits of its units while using up the possibly smallest amount of resources (costs). The essential question we address is the following: is there any relatively simple mechanism based on obvious assumptions and still leading to a complex hierarchical network observed in most societies made up of predominantly selfish (inherently non-cooperating) individuals/actors? Such a network would suggests the existence of a kind of cooperation, or at least a situation in which the individuals feel interested to occupy their varying positions.

First of all, it can be argued that most of the problem solving situations involve an estimation of the right answer[24]. Our assumptions take into account that i) the groups of individuals are typically embedded into a changing environment and better adaptation to the changes (finding out about the new state of the environment as soon as possible) is one of the core advantages an individual or, alternatively, the whole group can have. Importantly, ii) the abilities of the individuals to gain advantage from their environment on their own is obviously diverse, thus, iii) individuals are trying to follow the decisions of their group mates (learn from them) in proportion with the degree they trust the level of judgment of the other actors as compared to their own level of competence. iv) Maintaining a decision-making connection with a group mate has a cost (effort). When these common and natural assumptions are integrated into our model, the process results in the emergence of a collaboration-like structure in which the leader-follower relationships manifest themselves in the form of a multi-level, directed hierarchical network. Neglecting any of the above four points leads to loosing the emergence of a multi-level hierarchical structure.

Before specifying the rules of the model, we define a changing environment (the state of which the individuals have to guess to gain benefit) that is as simple as possible, while still varying in an unpredictable way. The state of the environment is chosen to have one of the $l$ (1,2 ... $l$), where $l$ is the number of states the environment can have. A given state is randomly replaced with a randomly chosen other one with a probability $p$, thus, the characteristic time between flips is about $1/p$ steps during which the units can "learn" about the environment's actual value.

The main steps of our basic model are (see also Fig. 1):

1. Each individual has a pre-defined ability or fitness $a_i$ (crucially: not *a priori* known by the other individuals[25]) to make a proper guess (with probability $a_i$) of the state of the environment. Their actual guess in each turn of the iterative process of building up a network is based on trust (nominations and choice from the other's decisions based on the trust matrix, see below) by making a weighted average of the decisions of the most trusted $k=1,2\ ..n$ friends/colleagues/players and his/her own estimate (for details of 1. 2. and 3. see the Supplementary Information (SI)). In addition, the number of times the decision of a given individual $i$ can be copied in a time step is constrained so that it lies within a pre-defined interval (typically from 2 to 7) since maintaining connections by $i$ has a cost.

2. After everyone completed a round of making their guesses/estimates, the actual state of the environment is revealed, letting the units learn which ones of them have made the right guess/decision.

3. The above information allows the construction of an updated trust matrix $T$. The $ij$-th element, $T_{ij}$, of the trust matrix is proportional to the number of times individual $i$ made use of the decision of individual $j$ (acceptance) in such a way that the decision by $j$ contributed positively to the correct guess.

The above process goes on iteratively and typically converges to a trust matrix having values depending on a non-trivial way on the original $a_i$ values due to the stochastic non-linear dynamics defined above (propagation, decision and feedback, see Fig.1). Obviously, this simplest variant (1-3) can be easily modified to take into account additional common factors influencing the decisions of the individuals.

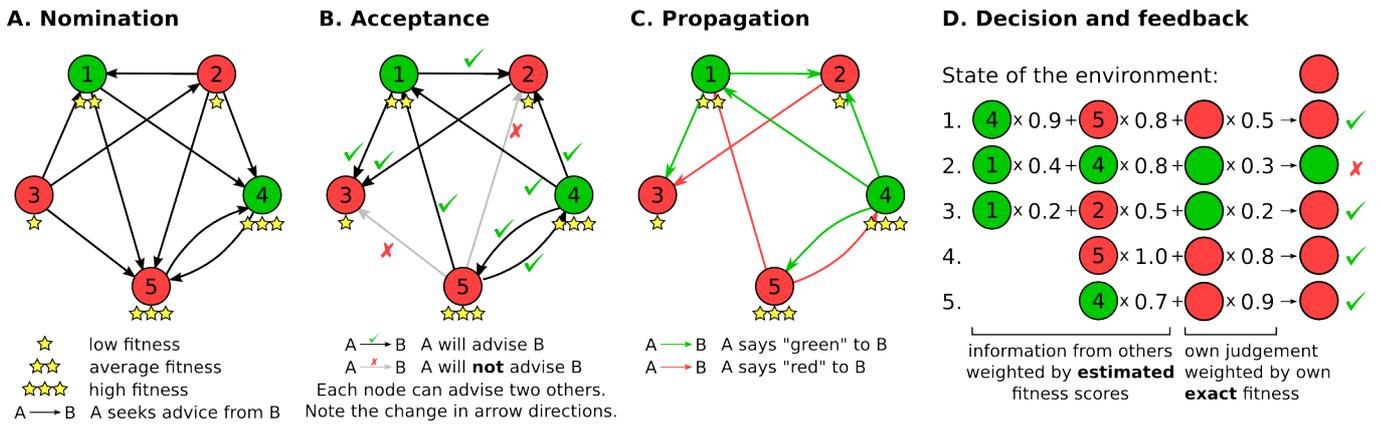

**Figure 1: Schematic explanation of the main stages of a single step in our model.** The environment is assumed to have two states: green and red. Nodes of the graph represent the individuals in the model and they are coloured according to their response in the last round (green or red). In the nomination phase, an edge from node A to node B means that A intends to seek advice from B. In the acceptance phase, the directions of the edges are reversed: a black edge from A to B with a tick mark means that A will advise B; a gray edge with a cross means that A refused to advise B in this round. In the propagation phase, edges represent the direction of information propagation and their colour is equal to the last response (i.e. colour) of the source node. The rightmost panel describes how the new responses of the individuals are derived from their own judgment and the information they received from others.

Thus, the main outcome of our model is an evolving and converging trust matrix expressing the level/strength of the social ties emerging as a result of the multiple interactions aimed at adapting to the changing environment as efficiently as possible. In this approach the individuals are optimizing their behaviour by simply copying decisions from the better performing ones. It is this trust matrix which – after appropriate evaluation and visualization – possesses the information about the hierarchical nature of the collective decision making process. A typical run starts with a uniform (except the diagonal) trust matrix which then evolves in time in such a way that after some time, the permanently changing $T$ more or less suddenly jumps into a state which optimizes the overall performance of the group to a much higher degree than a random matrix.

If we assume that the network representing the trust matrix contains its most relevant elements (the strongest ties between the actors), then using a few natural criteria the corresponding network can be derived. Once a graph representation of the trust matrix is obtained, we can proceed to defining quantities characterizing the extent of hierarchy present in the graph. We will use two complementary measures: the normalized fraction of forward arcs[26] and the so-called global reaching centrality (GRC) as defined in Ref. 27.

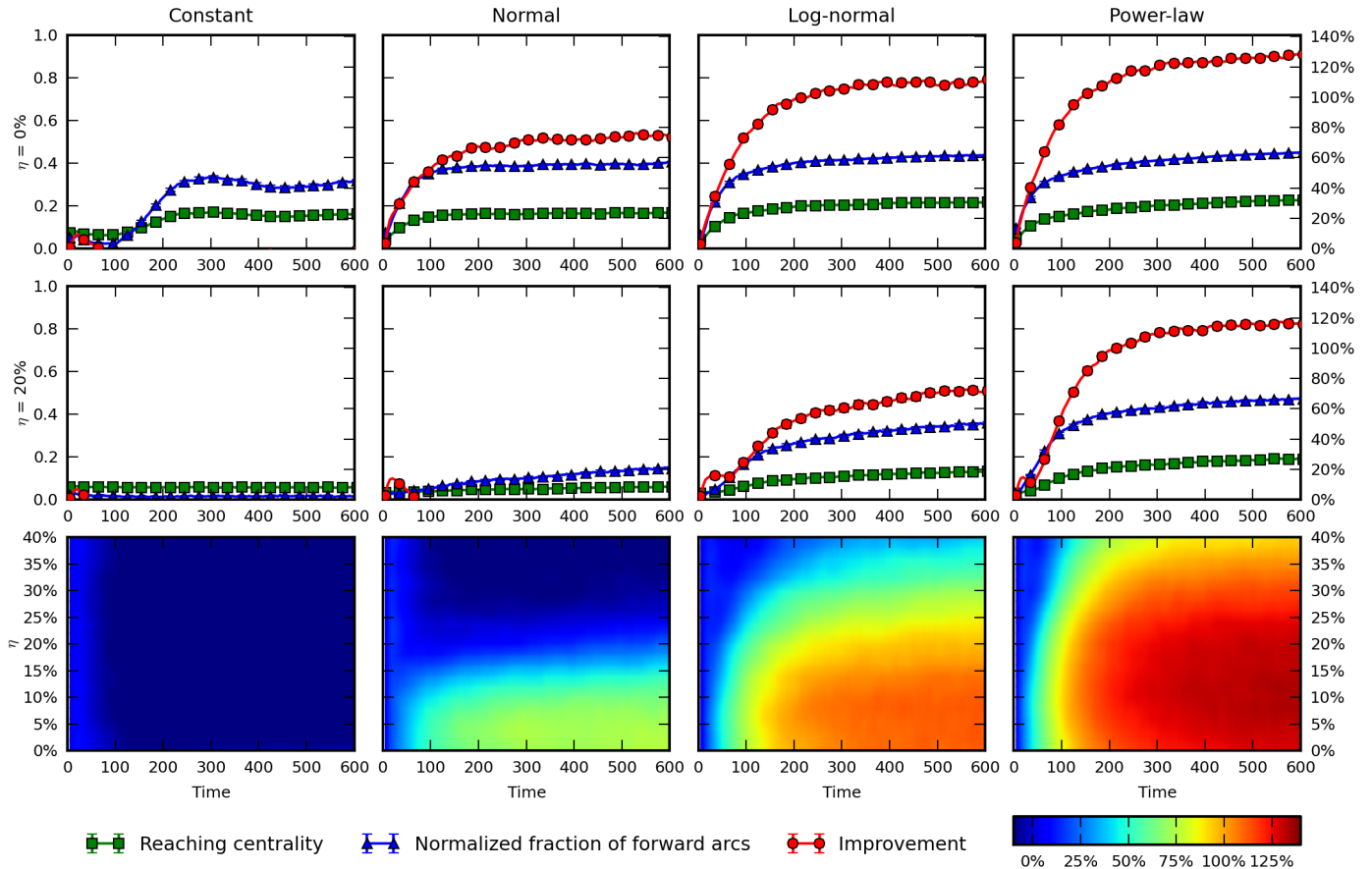

**Figure 2: Behaviour and performance of our model as a function of time and noise for various ability distributions**. The columns correspond to constant, normal, log-normal and power-law fitness distributions with a mean fitness of 0.25 and a variance of 1/48. The upper row corresponds to the case of no noise; the middle row corresponds to 20% relative noise. The green and blue lines correspond to two hierarchy measures (fraction of forward arcs[26] and global reaching centrality[27], expressed as numbers between 0 (no hierarchy) to 1 (maximal hierarchy that is theoretically possible). Red lines indicate the improvement of the overall performance of the individuals, expressed as percentages on the right axis. The heat maps in the bottom line represent the improvement as a function of time and relative noise level; note that the log-normal and power-law distributions are more tolerant to noise than the normal ability distribution (for more detailed definitions of the above quantities see the SI). Each data point is averaged from 500 trials with $N$=256 individuals; error bars represent the standard error of the mean but they are smaller than the corresponding markers on the plot. A smaller scale run of the model is visualized by the Supplementary Video 1.

The simulations we present here were conducted on groups of 256 agents. The model was run for 1000 time steps, and each data point on the plots was averaged from 500 simulations. We have assumed no noise (i.e. $\eta = 0$). The results for different classes of ability (fitness) distributions are presented in Fig. 2. For each distribution class we have studied (constant, normal, log-normal and Pareto, i.e., algebraic), the mean of the distribution was set to 0.25 and the standard deviation was set to sqrt(1/48) - the latter was chosen because this is the standard deviation of the uniform distribution on the [0; 0.5] interval. The results show a strong tendency of the model to promote hierarchical organization among the agents. However, there are some remarkable differences resulting from the various distributions we considered. Noise was introduced through adding a Gaussian distributed random value at the nomination phase (Fig. 1) to the perceived ability scores with a standard deviation expressed in percents of the average fitness.

We also plotted the cumulative performance improvement gained by the agents from copying decisions. An abbreviated version of the definition of the performance improvement of a given agent (for details see the SI) involves the average number of its good guesses taken into account with an exponentially decaying weight (backward from the given time step and with a "half-time" of 50 steps). Then, to obtain a relative improvement, the average ability is subtracted from this "score" and the result is divided by the average ability. We find that fat-tailed distributions not only promote the emergence of hierarchy in the model but also enable the actors to reach a consensus state where the vast majority copies the decision of a single leader in the largest component. In contrast, constant and normal distributions give rise to many competing leaders, each with their own subset of followers, and the different leaders do not communicate with each other. The performance improvement score shows significant (in cases well over 100%) improvement for the power law distribution as compared to, e.g., the constant distribution that results in no measurable improvement.

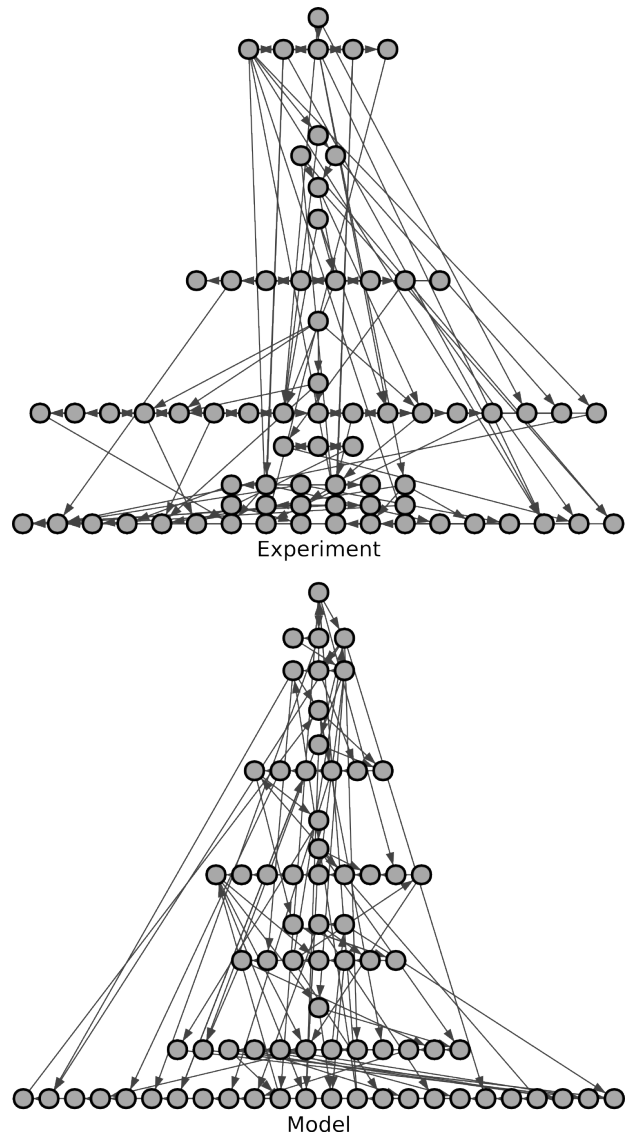

**Figure 3: Qualitative comparison of the experimental and the modelling results.** The bottom network was generated by our approach showing features similar to those obtained during the Liskaland experiment. The data were plotted using the method introduced in Ref. 27.

Naturally, our approach is relevant to the better understanding of the existing complex systems if its results agree qualitatively, or, ideally, quantitatively with those observed in nature or society. Here we shall take two approaches. In the first one we show that the main parameters of the networks we obtain are very similar to a selection of real systems. Table 2 in the Supplementary Information demonstrates that indeed this is the case for a variety of real-life networks described in the literature. Further significant evidence in favour of our approach comes from an experiment which has recently been carried out aiming at observing how the leadership-followership relations are built up among the members of a group of 86 people. In the setting of the experiment the human actors are participating in a process in which they are interested in gaining advantage (making a larger amount of virtual money) during a camp organized along the economic theory of Liska[9]. During the experiment the participants were interested in getting good advice from other participants and the information about their tendencies to follow others were recorded using an on-line questionnaire (for details see Refs. 9. and 10.). Fig. 3 shows the outcome of the experimentally registered network of directed interactions between the human participants built up during their one week long interactions as compared to one of the typical hierarchical networks our model predicts, for the same parameters as those of the experimentally observed network, i.e., 73 nodes (13 participants remained segregated) and 142 edges. The good qualitative and quantitative agreement between the experimental and the model network provides a promising evidence in favour of our approach.

One of our main propositions drawn from the numerical experiments is concerned with the effect of perturbations on the kind of hierarchical structures we obtained. To get an insight into this important question we carried out the following simulation of the model. We let the emerging structure converge to a meta-stable state, and then, after 500 time steps increased the relative amount of noise to 40% of the mean of the ability values. The results are illustrated on Fig. 4.

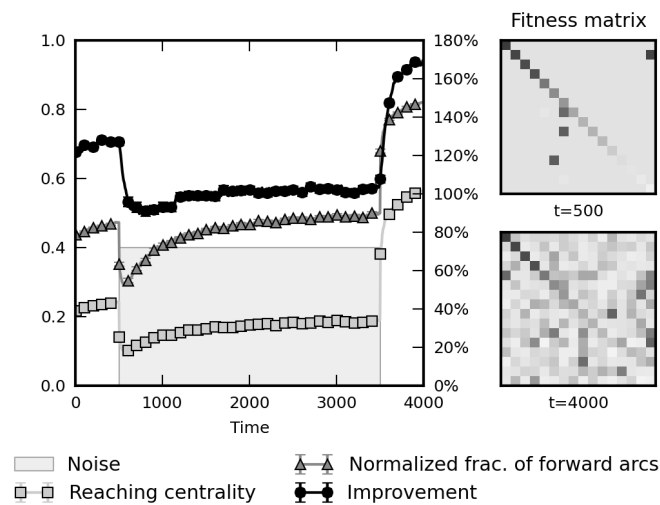

Figure 4: **The effect of transient noise on the structure of the model and the fitness matrix**. $t=0$ represents a starting configuration where the improvement has already converged to a stationary level in the absence of noise. A relative noise of 40% was added to the model between $t=500$ and $t=3500$. The panels on the right show a 16x16 sub-matrix of the fitness matrix at $t=500$ (i.e. before the noise was turned on) and at $t=4000$ (i.e. after the noise was turned off). Note how the performance improvement increased at $t > 3500$ compared to the baseline level between $t=0$ and $t=500$, and that the estimates in the fitness matrix became more diverse since the agents were forced to experiment with the structure of the communication graph due to the high level of noise. A small-scale simulation of the above kind is visualized by the Supplementary Video 2.

As expected, introducing relatively large perturbations resulted in a strong restructuring of the trust matrix leading to a sharp decrease of the improvement value. After this, with time, the system was gradually improving, in a way adapting to the larger magnitude of the noise. Finally, when the noise was completely switched off at time step 3500, the simulations showed a very fast growth in the improvement accompanied with a sharply increasing values of the hierarchy measures.

In conclusion, intuition would suggest the optimal network would be a very simple 2 level hierarchy in which all of the actors would copy the estimates of the best individual (the one with the highest $a_i$). However, this is not what happens and there are good reasons for that. First of all, due to the stochastic nature of making decisions, some of the units with a smaller ability may guess the right state in a given round (thus, building up trust in them and placing them on a higher level of the hierarchy), while an individual with a higher ability may fail to do so. Second, since the individuals with the best abilities have a relatively small number of maximal contacts, there is a constraint leading to a "chain" of interactions through which agents with smaller abilities can connect themselves to the best performing ones.

Perhaps the most remarkable feature of the hierarchical structures we obtain is their stability against perturbations. Once a "locally" optimal network is obtained (due to the mechanism of copying decisions with the above mentioned aspects) it has a strong tendency to stay on, since its restructuring to a more efficient network would involve going through configurations which are much less efficient. This is rather reminiscent to those occurring in many systems described in terms of statistical mechanics, for example configurations into which the so called spin glasses[28] freeze into at low temperatures. Thus our generic model quantitatively exhibits the two main and non-trivial (because acting in opposite directions) features of complex systems (such as, e.g., organisms or organizations) which are i) their capability to adopt to the changing environment while ii) still maintaining their essential structure (adaptation versus stability). These features show up most remarkably in the case of highly skewed ability distributions.

**Methods Summary**

**Model description**

Our model contains $n$ independent individuals embedded in a $l$-state discrete environment. The simulation consists of steps, and within each step, individuals must infer the current state of the environment. State changes occur between steps with probability $p$. The fitness of individual $i$ (i.e.

the probability of inferring the state on its own) is denoted by $a_i$. Each individual knows its own fitness but not the fitness of others.

Individuals are allowed to seek advice from their peers, and the final decision of each individual is based on a weighted average of the individual's own guess and the information received from others. Individuals strive to select partners with high fitness scores and therefore estimate the fitness of others based on the number of interactions between them and the number of times the information exchanged was correct. The number of contacts an individual tries to establish in a single round inversely depends on its own fitness. At the same time, the number of contact requests individual $i$ can handle is limited by its link capacity $c_i$.

**Measuring group performance and hierarchy**

Individuals get a score of 1 per round when responding correctly and zero when they respond incorrectly. The performance of an individual is the exponential moving average of its scores with a half-life of 50 steps, and the performance of the group is the average performance of its members. The group performance score is compared to the baseline performance with no communication – this is equal to the average fitness within the group. The relative improvement of the group is defined as the group performance divided by the average fitness score, minus one.

We also define the communication network as a snapshot of the interactions in a given round of the model where an edge points from individual $i$ to $j$ if $i$ provided information to $j$ in that round. The extent of hierarchy in the network is evaluated by the global reaching centrality[27] and the normalized size of the largest cycle-free arc set[26].

**Supplementary Information** accompanies the paper.

**Acknowledgements**. We are grateful to K Ozogány for helping to produce some of the illustrations. This research was partially supported by the EU ERC COLLMOT project.

**Author Contributions.** T.N. and T.V. designed the model. T.N. wrote the program and carried out the simulations. T.N. and T.V. wrote the paper.

**Author Information.** Correspondence and requests for materials should be addressed to T. V. (vicsek@hal.elte.hu).

Hierarchical self-organization of non-cooperating individuals

Supplementary Information

T. Nepusz and T. Vicsek

## Model description

Our model contains *n* independent individuals that are embedded in an environment which is always in exactly one of *k* possible states. The simulation consists of rounds and the state of the environment is constant within a round but may change states between rounds with probability *p*. When a state change occurs, the new state is always selected uniformly randomly from the set of possible states minus the current state of the environment.

The goal of the individuals in each round of the simulation is to find out the current state of the environment. Individuals that "guess" or "infer" the state of the environment accurately gain positive feedback (e.g., a score of 1) at the end of each round, while those that do not manage to infer the state of the environment receive negative feedback (e.g., get a score of zero). The total score of an individual is an exponentially moving average of all the feedback received during the simulation, with a half-life of 50 rounds. In other words, the weight of feedback diminishes exponentially over time in a way that feedback received 50 rounds ago has weight 0.5, feedback received 100 rounds ago has weight 0.25 and so on.

The individuals in the simulation differ in their ability of inferring the state of the environment correctly without any external input. The probability of individual *i* responding correctly in a round on its own is denoted by $a_i$, which we will call the *ability* of the individual from now on. In other words, when an individual is forced to decide on its own, it chooses the current state of the environment with probability $a_i$ or chooses a different state uniformly at random with probability $1-a_i$. It is important to note that each individual is aware of its own ability but does not know the ability of others *a priori*.

The hierarchical structure emerges from the exchange of information between the individuals. The exact mechanism of information exchange will be described later. Here we only mention that individuals are allowed to seek advice from others and take their guesses into account when making their own decision, and they strive to select partners with high ability scores. Since the exact ability scores are not publicly available, each individual maintains a vector of estimated ability scores of others. In particular, let $t_{ij}$ denote the ability score of individual *j* as perceived (or estimated) by individual *i*. $t_{ij}$ depends on two variables: $n_{ij}$, the number of rounds in which individual *j* passed information to individual *i*, and $s_{ij}$, the number of rounds in which the information that individual *j* passed to individual *i* turned out to be correct. $t_{ij}$ is then calculated as follows:

$$t_{ij} = \frac{s_{ij}+1}{n_{ij}+k}$$

where the constants in the numerator and denominator serve the purpose of providing a meaningful estimate even in the absence of interaction between individuals $i$ and $j$ (i.e. when $n_{ij}$ is zero). The above formula is consistent with the Laplace-Bayes estimator for $k$ possible outcomes. We assume that $t_{ii}$ is equal to $a_i$, i.e. each individual knows its ability score exactly.

Finally, we introduce limitations on the number of contacts an individual is allowed to establish within a single round. Since the propagation of information is asymmetric (i.e. individual $i$ may not necessarily reveal its own opinion to individual $j$ when he asks $j$'s opinion), we impose constraints both on the number of other individuals a given person will try to contact in a round and on the number of other individuals a given person can provide information to; the latter will be referred to as the *link capacity* $c_i$ for individual $i$. The number of contacts an individual tries to establish depends on its own ability: highly fit individuals will try to establish a smaller number of contacts since they confide in their own opinion, while individuals with a lower ability will compensate by asking more people around them. The link capacities are drawn from a uniform distribution to model the fact that some people are more social and/or have better "management" or "organizational" skills than others. Table summarizes the concepts and notations we have introduced so far.

| Notation | Description |
| --- | --- |
| $n$ | The number of individuals |
| $l$ | The number of states in the environment |
| $p$ | Probability of state change |
| $a_1, a_2, …, a_n$ | The ability scores of the individuals |
| $c_1, c_2, …, c_n$ | The link capacities of the individuals |
| $k$ | The maximum number of peers that an individual may ask in a single round |
| $t_{ij}$ | The ability of individual $j$ as perceived by $i$ |
| $n_{ij}$ | Number of rounds in which individual $i$ asked $j$ |
| $s_{ij}$ | Number of rounds in which individual $i$ received correct information from individual $j$ |

*Table 1: Parameters and notations used in the model description*

Each round of the game consists of five phases as follows:

1. In the *nomination* phase, each individual assembles a list of individuals they wish to receive information from in the current round. In particular, an individual with ability $a_i$ will nominate $k(1-a_i)$ partners (rounded up to the nearest integer), where $k$ denotes the maximum number of individuals that one can ask in a single round. This formula essentially scales the number of partners linearly from 1 to $m$ as the ability decreases from 1 to 0. The partners are nominated based on their perceived ability: individual *i* will nominate those that he assumes to be the fittest based on his own experience. Ties in perceived ability scores are broken randomly. Individuals then nominate an additional $k(1-a_i)$ peers who will act as a backup in case someone rejects the contact request in the next phase due to link capacity constraints.

2. In the *acceptance* phase, each individual accepts at most $c_i$ partners from those who wish to contact him, preferring those with whom he has also been in contact in the previous round. The remaining partners that nominated the individual will be rejected and no information will be propagated to them in this round.

3. In the *propagation* phase, everyone sends their guess from the *previous* round to all the communication partners they have accepted.

4. In the *decision* phase, everyone calculates the majority opinion based on the information they received and their own guess with respect to the state of the *current* round. Pieces of information from others are weighted by the estimated ability scores of the partners the information came from, while the own opinion is weighted by the actual ability score. We wish to stress that the information received from others is based on the *previous* round, meaning that there is a penalty associated to relying too much on others since the state of the environment might have changed in the current round.

5. In the *feedback* phase, the current state of the environment is revealed to everyone. Each individual then re-calculates the perceived ability of others by updating the $n_{ij}$ and $s_{ij}$ counters: $n_{ij}$ is increased by 1 for every individual *j* that *i* received information from, while $s_{ij}$ is increased by 1 for every individual *j* whose information turned out to be correct.

The phases described above are then iterated for a number of rounds, allowing the matrix of perceived ability scores and also the communication network to evolve into a configuration where the overall performance of the group is higher than the baseline performance that we would have observed if we allowed no communication. The performance measures and the structural properties of the communication network will be described in the next section.

In the manuscript, we describe several simulations where the behaviour of the individuals was perturbed by random noise. Noise was injected in the nomination phase (see item 1 above) where we added Gaussian noise with zero expected value and a given standard deviation to the perceived ability scores. In order to make the amount of noise comparable with the average ability of the individuals, we express the standard deviation of the noise as a percentage relative to the average ability, and we refer to this quantity as *relative noise*. A relative noise of 100% means that the standard deviation of the noise component is exactly as large as the average ability of the individuals.

## Measuring group performance and the extent of hierarchy

The overall performance of the group depends on the number of times the members of the group managed to infer the state of the environment correctly. Since each member responds on its own, we will simply calculate individual performance scores and define the overall performance of the group as the average performance of its members.

Let us assume that individuals responding correctly in a round are awarded a score of 1, while incorrect answers result in a score of zero. The performance of an individual is then simply defined as the exponential moving average of its entire score history, where the weight of the most recent data point is 1 and the weights of past data points decay in an exponential manner. It is common to define the decay factor in terms of a half-life constant λ, meaning that the weight of a feedback received λ steps ago is exactly 1/2; the weight of a feedback received 2λ steps ago is 1/4 and so on. In our simulations, λ was chosen to be 50. To avoid transients in the early stages of the simulation when the score history is significantly less than the half-life of the exponential moving average, we assume that the score history is preceded by an infinite sequence of $a_i$ values, representing the fact that the individual would have received a feedback of $a_i$ per round on average if we ran the simulation for a long time without allowing communication.

This latter observation also allows us to define a baseline to which we can compare the group performance. If we disallowed communication, member $i$ of the group would have obtained a performance score of $a_i$ after an infinite number of steps since he succeeds in inferring the state of the environment on its own with probability $a_i$. The baseline performance of the group can then be defined as the average ability score within the group. Finally, we can define the *relative improvement* of the group performance as the group performance divided by the average ability score minus one. A relative improvement of zero thus means that the group performs exactly as well as a hypothetical group without communication, a relative improvement of 1 (or 100%) means that the group performs twice as well as the baseline group, and so on.

Besides measuring the overall performance of the group, we are also interested in measuring whether there are signs of a hierarchical organization in the communication network among the individuals. The communication network is simply defined as a snapshot of the interactions in a given round of the game; the nodes of the networks represent the individuals, and an edge points from individual $i$ to $j$ in the network representation if $i$ provided information to $j$ in the round being examined. The extent of hierarchy present in the network is evaluated by two independent means: the *global reaching centrality* measure defined by Mones et al (2012) and the normalized size of the *largest cycle-free arc set*.

The global reaching centrality of a network is measured as follows. Let us define the *local reaching centrality* $C_R(i)$ of node $i$ in the network as the number of other nodes that are reachable from it via directed paths divided by $n$-1, where $n$ is the number of nodes in the network. Furthermore, let $C_R^{max}(i)$ denote the maximal local reaching centrality in the network. The global reaching centrality *GRC* is then defined as follows:

$$GRC = \frac{\sum_{i=1}^{n}\left[C_R^{max} - C_R(i)\right]}{n-1}$$

The value of the global reaching centrality of a network is high when the maximal local reaching centrality is significantly different from the local reaching centrality of most of the nodes; in other words, if there are only a few nodes from where the whole network can be reached while most other nodes can reach only a small part of the network. GRC scores are strictly non-negative and the maximal GRC value of 1 is attained for star graphs. Regular and irregular trees also have a high GRC score, while Erdős-Rényi random networks attain a low GRC score of around 0.058 ± 0.005 (Mones et

al, 2012).

An alternative hierarchy measure we consider is the normalized size of the *largest cycle-free arc set*. Consider an arbitrary total ordering of the nodes of a network. An edge of the network is called a *forward arc* with respect to the ordering if the edge points from a node with smaller rank towards a node with larger rank, otherwise the edge is called a *feedback arc*. Note that the sub-graph that consists of forward arcs or feedback arcs only is cycle-free by definition. The *largest cycle-free arc set* is thus the set of forward arcs in an ordering that yields the largest such set. The usage of the largest cycle-free arc set as a proxy for measuring the hierarchy is justified by the intuition that a network is more hierarchical if we need to remove a smaller number of edges to make it cycle-free.

The size of the largest cycle-free arc set on its own is an un-normalized measure in the sense that nodes of completely random Erdős-Rényi networks (which have no built-in hierarchy by definition) can still exhibit a relatively large cycle-free arc set, especially if the network is sparse and consists of many disconnected components. To this end, we measure the *normalized size of the largest cycle-free arc set* instead. For a network with $n$ nodes and an average degree of $d$, the normalization is done by subtracting the expected size of the largest cycle-free arc set in an Erdős-Rényi network with the same number of nodes and average degree from the observed size of the largest cycle-free arc set, and dividing it by the number of edges minus the expected size. This normalization yields a score with a theoretical maximum of 1; furthermore, the normalized score will be zero if the network contains exactly as many forward arcs as an Erdős-Rényi network with the same number of nodes and edges.

Calculating the exact size of the largest cycle-free arc set is NP-hard, especially for $k$-edge-connected graphs where each edge participates in many cycles. However, there exist heuristics that approximate the size of the largest forward arc set (and thus the largest cycle-free arc set) accurately in sparse graphs (Eades et al 1993). In our study, we used the following simple approximation scheme (Eades et al, 1993) that runs in linear time with respect to the number of edges in the graph:

1. Construct two empty queues of nodes, which we call the *front queue* and the *back queue*.
2. Find all the nodes with zero in-degree, append them to the front queue and remove them from the graph. Repeat this step until no such nodes can be removed.
3. Find all the nodes with zero out-degree, reorder them to the back queue and remove them from the graph. Repeat this step until no such nodes can be removed.
4. If there is at least one node left in the graph, find the node for which the difference between its out-degree and in-degree is the largest, append it to the front queue, remove it from the graph and continue from step 2.
5. Append the back queue to the front queue to obtain the final ordering. The approximated forward arc set can then be derived from the ordering.

The expected size of the largest cycle-free arc set in Erdős-Rényi networks was determined with simulations. For each required value of the network density $d$, we generated 1000 Erdős-Rényi networks, searched the largest forward arc set with the above heuristic, and averaged the size of these sets over all the generated instances.

**Model parameters to generate real-world-like networks**

In this section, we present several real-world networks and describe them with three key parameters of interest: the number of nodes N, the average degree <k> and the global reaching centrality GRC (Mones et al, 2012), the latter serving as a quantification of the extent of hierarchy present in the network. We claim that it is possible to set the parameters of our model in a way that the communication network between the individuals yields similar average node degrees and a similar GRC measure. Table 2 of this Supplementary Information contains the key parameters of the real-world networks we have studied and the values of these parameters obtained from an appropriate parameterization of the model. Table3 contains the exact parameter settings we used to obtain our results with the model. The general guidelines to set the parameters were as follows:

- The number of individuals is equal to the number of nodes in the real-world network.
- The environment has 5 states and a state change occurs with probability 0.1 between steps.
- The individual ability scores are Pareto-distributed with mean $\sigma$ (given in Table 3) and a variance of 1/48.
- Link capacities are uniformly distributed between $c_{low}$ and $c_{high}$ (given in Table 3).
- The maximum number of peers that an individual may ask in a single round are given by $k$ in Table 3.
- The relative noise added to the perceived ability scores is given by $\zeta$ in Table 3.

| Network | N | <k>$_{real}$ | GRC$_{real}$ | <k>$_{model}$ | GRC$_{model}$ |
|---|---|---|---|---|---|
| Prison[3] | 67 | 2.716 | 0.172 | 2.617 ± 0.14 | 0.181 ± 0.11 |
| Liskaland[4] | 71 | 1.971 | 0.308 | 2.060 ± 0.10 | 0.288 ± 0.09 |
| TRN-Yeast[5] | 688 | 1.568 | 0.116 | 1.672 ± 0.11 | 0.158 ± 0.07 |
| C.elegans[6] | 1173 | 2.442 | 0.048 | 2.570 ± 0.18 | 0.095 ± 0.02 |

*Table 2: Comparison of real-world networks with synthetic networks generated by our model. The number of nodes N was identical for both networks. <k>$_{real}$ and <k>$_{model}$ are the average degrees in the real network and the synthetic networks. GRC$_{real}$ and GRC$_{model}$ are the global reaching centrality (Mones et al, 2012) of the real and the synthetic networks. Values for the synthetic networks were averaged from at least 20 instances at t=1000 and are given as mean ± standard deviation.*

| Network | $\sigma$ | $c_{low}$ | $c_{high}$ | $k$ | $\zeta$ |
|---|---|---|---|---|---|
| Prison | 0.25 | 3 | 10 | 3 | 40% |
| Liskaland | 0.4 | 1 | 7 | 3 | 30% |
| TRN-Yeast | 0.25 | 1 | 3 | 2 | 10% |
| C.elegans | 0.35 | 3 | 15 | 3 | 40% |

*Table 3: Model parameters that generate networks with average degrees and global reaching centrality (GRC) scores similar to real networks.*

## Supplementary video clips

**Supplementary video 1**. Starting from a random network of 73 agents, the rule of copying the decisions of the most trusted neighbours results in a relatively quickly converging hierarchical network having a larger overall score (better performance) than the average of the abilities of the individuals.

**Supplementary video 2**. The ties among the 32 agents self-organize into a hierarchical network and freeze into a local maximum during the first third of the run. Next, an "annealing" is applied (the noise term is considerably increased). Finally, the level of noise is set back to zero and a new maximum is found which is both more stable and corresponds to a larger overall score